\documentclass[aps,prd,twocolumn,superscriptaddress,showpacs,letter,nofootinbib]{revtex4-1} 
\usepackage[utf8]{inputenc}
\usepackage{graphicx}
\usepackage{dcolumn}
\usepackage{bm}
\usepackage{cancel}
\usepackage{color}             
\usepackage{epsfig}
\usepackage{amssymb}
\usepackage{amsmath}
\usepackage{multirow}
\usepackage{hyperref}
\usepackage{cleveref}
\usepackage{lineno,color}
\usepackage{epstopdf}

\begin{document}

\title{Odderon in the light of collider low-$t$  data}
\thanks{Presented at ``Diffraction and Low-$x$ 2024'', Trabia (Palermo, Italy), September 8-14, 2024.}

\author{E.~G.~S.~Luna}
\email{luna@if.ufrgs.br}
\affiliation{Instituto de F\'isica, Universidade Federal do Rio Grande do Sul, Caixa Postal 15051, 91501-970, Porto Alegre, Rio Grande do Sul, Brazil}
\author{M.~G.~Ryskin}
\email{ryskin@thd.pnpi.spb.ru}
\affiliation{Petersburg Nuclear Physics Institute, NRC Kurchatov Institute, \\Gatchina, St.~Petersburg, 188300, Russia}
\author{V.~A.~Khoze}
\email{v.a.khoze@durham.ac.uk}
\affiliation{Institute for Particle Physics Phenomenology, University of Durham, \\ Durham, DH1 3LE, UK}
 

\begin{abstract}

Odderon is the $C$-odd amplitude that does not fall (or decrease very slowly) with energy. 
The expected amplitude is small and mainly real. Therefore, extracting it from the data on top of a much larger $C$-even contribution is challenging. The only chance is to consider the very low $|t|$ region of Coulomb nuclear interference or the diffractive dip region.
 Here we perform the analysis of elastic scattering $pp$ and $\bar pp$ data at low momentum transfer $|t| < 0.1$
GeV$^2$ within large collider energy interval $\sqrt s = 50$ GeV $-$ 13 TeV in order to evaluate quantitatively
the possible Odderon contribution. We use the two-channel eikonal model, which naturally accounts for the screening of the Odderon amplitude by the $C$-even (Pomeron) exchanges.  

\end{abstract}


\maketitle

\section{Introduction}

Odderon is the $C$-odd amplitude which does not fall (or die very slowly) with energy. Theoretically, there is no reason {\em not}  to have such an amplitude. Moreover, it appears in perturbative QCD where in the lowest $\alpha_s$ order it is given by the three gluon exchange when all three gluons are symmetric in colour (i.e. convoluted by the colour SU(3) tensor $d^{abc}$).

Since the Odderon amplitude is expected to be rather small the best chance to observe it on top of a much larger $C$-even contribution is either in the diffractive dip region where the imaginary part of $C$-even amplitude vanishes or by measuring the real part of $pp$ ($\bar pp$) elastic amplitude. Recall that due to dispersion relations the real part of high energy $C$-even amplitude is relatively small (Re$A_{even} \ll$ Im$A_{even}$).

The real part of proton-proton amplitude can be measured via the Coulomb-nuclear interference (CNI) at very low momentum transferred $|t|$. In 2018 TOTEM claimed the Odderon discovery based on  the measurements of
the total cross-section and the real part of the forward
elastic pp amplitude at 13 TeV~\cite{rho-T}. The observed
value of the ratio of real to imaginary parts of the forward
scattering amplitude, namely $\rho = (0.09 - 0.10) \pm 0.01$
turned out to be noticeably smaller than the predicted
value ($\rho = 0.13 - 0.14$ \cite{COMP}) coming from dispersion relations for the case of pure $C$-even interactions.

This prompted a renewal of interest in the potential existence of the high-energy $C$-odd (Odderon) contribution.

The new ATLAS/ALFA data recently confirmed this
value of $\rho$ \cite{atl13}. However, the value of the total cross-section
at 13 TeV reported by the ATLAS/ALFA team, $\sigma_{tot} =
104.68 \pm  1.09$ mb~\cite{atl13}, is approximately 5\% lower than the
average of values determined by TOTEM ($\sigma_{tot} = 111.6\pm 3.4$ mb, $\sigma_{tot} = 109.5\pm 3.4$ mb, and $\sigma_{tot} = 110.3\pm 3.5$ mb)
indicating that a smaller value of the real part of the $C$-
even amplitude should be expected from the dispersion
relations.
The relatively small value of $\rho$ can be explained by
the admixture of the $C$-odd amplitude, which survives
at high LHC energies.

\section{Formalism}

In the recent paper \cite{LRK} the available low $|t|<0.1$ GeV$^2$ data at $50\ \mbox{GeV}<\sqrt s<13$ TeV were analyzed
 including both the TOTEM and ATLAS/ALFA results.
Since at all LHC energies the total cross section measured  
by ATLAS is systematically smaller than that claimed by TOTEM, in the fit we added (as free parameters) the corresponding normalization factors, $N_i$. That is, the $\chi^2$ was calculated as
\begin{equation}
\label{chi}
\chi^2=\sum_{ij}\frac{(N_ids^{th}_{ij}-ds^{exp}_{ij})^2}{(\delta^{rem}_{ij})^2} + \sum_i \frac{(1-N_i)^2}{\delta^2_i} \ ,
\end{equation}
where $i$ denotes the particular set of data, while $j$ denotes the point $t_j$ in this set of data; $ds^{th}$ is the theoretically calculated $d\sigma/dt$ cross-section (\ref{th}) while $ds^{exp}$ is the value measured at the same $ij$ point experimentally; $\delta_i$ is the normalization uncertainty of the given ($i$) set of data and $\delta^{rem}_{ij}$ is the remaining error at the point $ij$ calculated as $(\delta^{rem}_{ij})^2=\delta^2_{tot,ij}-(\delta_i\cdot d\sigma^{exp}_{ij})^2$. As a rule the value of $\delta^{rem}$ is dominantly the statistical error~\footnote{A similar approach was used in \cite{Sel}.}.

 
 Two-channel eikonal model
\begin{widetext}
 \begin{equation}
 \label{2eik}
 A(s,t) ~=~is  \int^{\infty}_{0} b\, db\, J_{0}(bq) 
 \left[ 1 -\frac{1}{4}\, e^{i(1+\gamma)^{2}\Omega(s,b)/2} -\frac{1}{2}\, e^{i(1-\gamma^{2})\Omega(s,b)/2} - \frac{1}{4}\, e^{i(1-\gamma)^{2}\Omega(s,b)/2} \right]
 \end{equation}
\end{widetext} 
was used where the opacity $\Omega(s,b)=\Omega_{Pomeron}(s,b)+\Omega_{Odd}(s,b)$ is given by the sum of the $C$-even/Pomeron and the Odd terms.

The opacity function $\Omega_{i}(s,b)$ is related to the bare nuclear amplitude $F^{N}_{i}(s,t)$ through the Fourier-Bessel transform
\begin{eqnarray}
\Omega_{i} (s,b) = \frac{2}{s}\int^{\infty}_{0} q\, dq\, J_{0}(bq)\, F^{N}_{i}(s,t) ,
\label{opacity001}
\end{eqnarray}
where $i = {\Bbb P}, {\Bbb O}$ represent the Pomeron and Odderon exchanges, respectively.

The single Pomeron contribution is given by
\begin{eqnarray}
F^{N}_{\Bbb P}(s,t) = \beta_{\Bbb P}^{2}(t)\, \eta_{\Bbb P}(t) \left( \frac{s}{s_{0}} \right)^{\alpha_{\Bbb P}(t)} ,
\end{eqnarray}
where $\eta_{\Bbb P}(t) = -e^{-i\frac{\pi}{2}\alpha_{\Bbb P}(t)}$ is the even signature factor,
\begin{eqnarray}
  \beta_{\Bbb P}(t) = \beta_{\Bbb P}(0) e^{\left( At+Bt^{2}+Ct^{3} \right)/2}
\end{eqnarray}
 is the elastic proton-Pomeron vertex, and
\begin{eqnarray}
\alpha_{\Bbb P}(t) = 1 + \epsilon + \alpha^{\prime}_{\Bbb P} t +  h(\pi\pi)
\end{eqnarray}
is the Pomeron trajectory with the pion loop
insertion $h(\pi\pi)$ \cite{AG}.
 
The Odderon contribution is given by
\begin{eqnarray}
F^{N}_{\Bbb O}(s,t) =  \beta_{\Bbb O}^{2}(t)\, \eta_{\Bbb O}(t) \left( \frac{s}{s_{0}} \right)^{\alpha_{\Bbb O}(t)} ,
\label{odd1}
\end{eqnarray}
where $\eta_{\Bbb O}(t) = -ie^{-i\frac{\pi}{2}\alpha_{\Bbb O}(t)}$ is the odd signature factor, $\beta_{\Bbb O}(t)) = \beta_{\Bbb O}(0) e^{Dt/2} $ is the elastic proton-Odderon vertex,
and we fixed the Odderon trajectory to its largest QCD value $\alpha_{\Bbb O}(t) = 1$ (see e.g. \cite{BLV,B,E}).\\

We accounted for the Coulomb nuclear interference
$ A^{C+N} = A^{N} + e^{i\alpha \phi(t)}A^{C}$ and the Bethe phase $\phi(t)$. Accordingly, the elastic differential cross-section reads
\begin{eqnarray}
\label{th}
\frac{d\sigma}{dt}(s,t) = \frac{\pi}{s^{2}}\, \left|  A^N(s,t) + e^{i\alpha \phi}A^{C}(s,t) \right|^{2} .
\end{eqnarray}

\section{Results}

We obtained a quite satisfactory fit with $\chi^2=560$ for 504 degrees of freedom, $\nu$; 
$\chi^2/\nu=1.11$.
 Neglecting the Odderon we get a much larger $\chi^2=726$. The quality of the description of  Coulomb-nuclear interference region at 13 TeV is shown in Fig.1, right, while the energy behaviour of $\sigma_{tot}$ and $\sigma_{el}$ - in Fig.1, left.
 
 The parameters corresponding to the Pomeron and the Odderon exchanges are given in Table I.  The normalization factors for ATLAS data turn out to be close to 1 ($N_7=1.015,\ N_8=1.003,\ N_{13}=1.009$, where index denotes the value of $\sqrt s$ in TeV), while for the TOTEM we get  $N_7=1.077,\ N_8=1.121$ and $N_{13}=1.15$.\\
 
 The slope of the Odderon coupling $\beta_{\Bbb O}\propto \exp(Dt/2)$ in the fit was fixed to be $D=A/2$. However, the results practically do not depend on this value. In particular, varying $D$ from $0.1A$ to $0.9A$ we get the same total cross section $\sigma_{tot}=105.1$ mb at 13 TeV and the $Re/Im$ ratio $\rho=0.112 - 0.110$. The only difference is the Odderon-proton coupling which for a small $D$ becomes larger ($\beta_{\Bbb O}=1.09\pm 0.24$ for $D=0.1A$) in order to provide the same Odderon contribution from a smaller impact parameters, $b$, where the screening caused by the Pomeron is stronger.

\section{Conclusion}
The main lessons about the Odderon learned from this study are:
\begin{itemize}
\item The description using the Odderon improves the fit (with the Odderon  $\chi^2$ becomes much smaller).
\item The sign of the Odderon amplitude needed to describe the very low $|t|$ data is opposite to that predicted by the perturbative QCD three-gluon exchange contribution \cite{DL,FK,Ro}.
\footnote{The problem can be solved assuming that the Odderon coupling $\beta_{\Bbb O}(t)$ vanishes (or strongly decreases) at $t=0$. In this case, the dominant $C$-odd contribution at $t=0$ comes from the Pomeron-Odderon cut and has the opposite sign. }
 \item The Odderon-proton coupling, $\beta_{\Bbb O}(0)$, is smaller than that for the Pomeron. Moreover, after accounting via the eikonal the screening of seed Odderon by the Pomerons, the final $C$-odd contribution to $\rho$ at 13 TeV becomes quite small,
$\delta\rho=(\rho^{\bar pp}-\rho^{pp})/2\leq 0.004$ -- i.e. 10 times smaller than that ($\delta\rho=0.04$) originally claimed by TOTEM. 
\end{itemize}

\section*{Acknowledgment}

This research was partially supported by the Conselho Nacional de Desenvolvimento Cient\'{\i}fico e Tecnol\'ogico under Grant No. 307189/2021-0.

\begin{figure*}
\label{f1}
\begin{center}
\vspace{-2cm}
\hspace{-1cm}
 \includegraphics[scale=0.39]{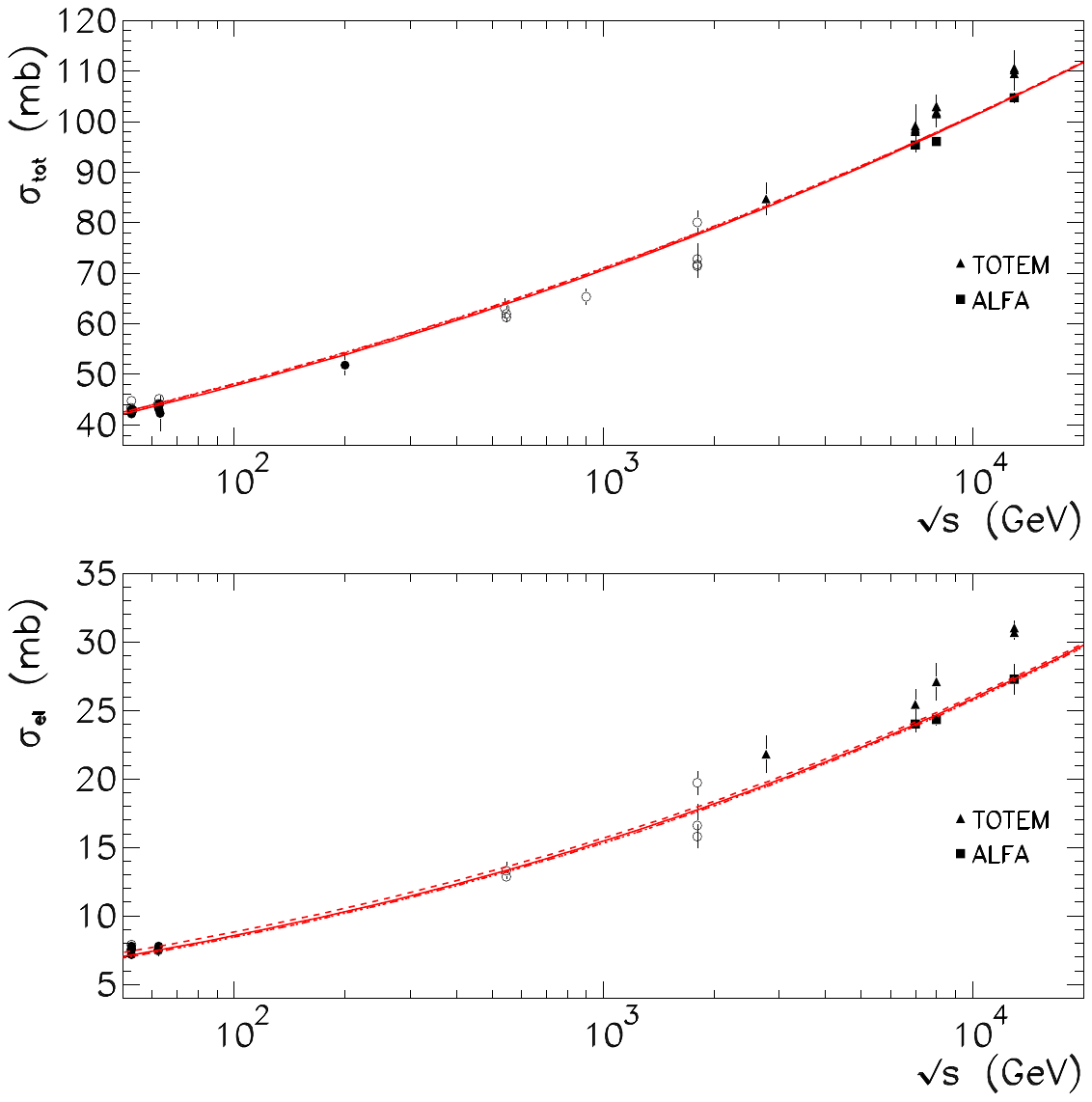}
 \includegraphics[scale=0.39]{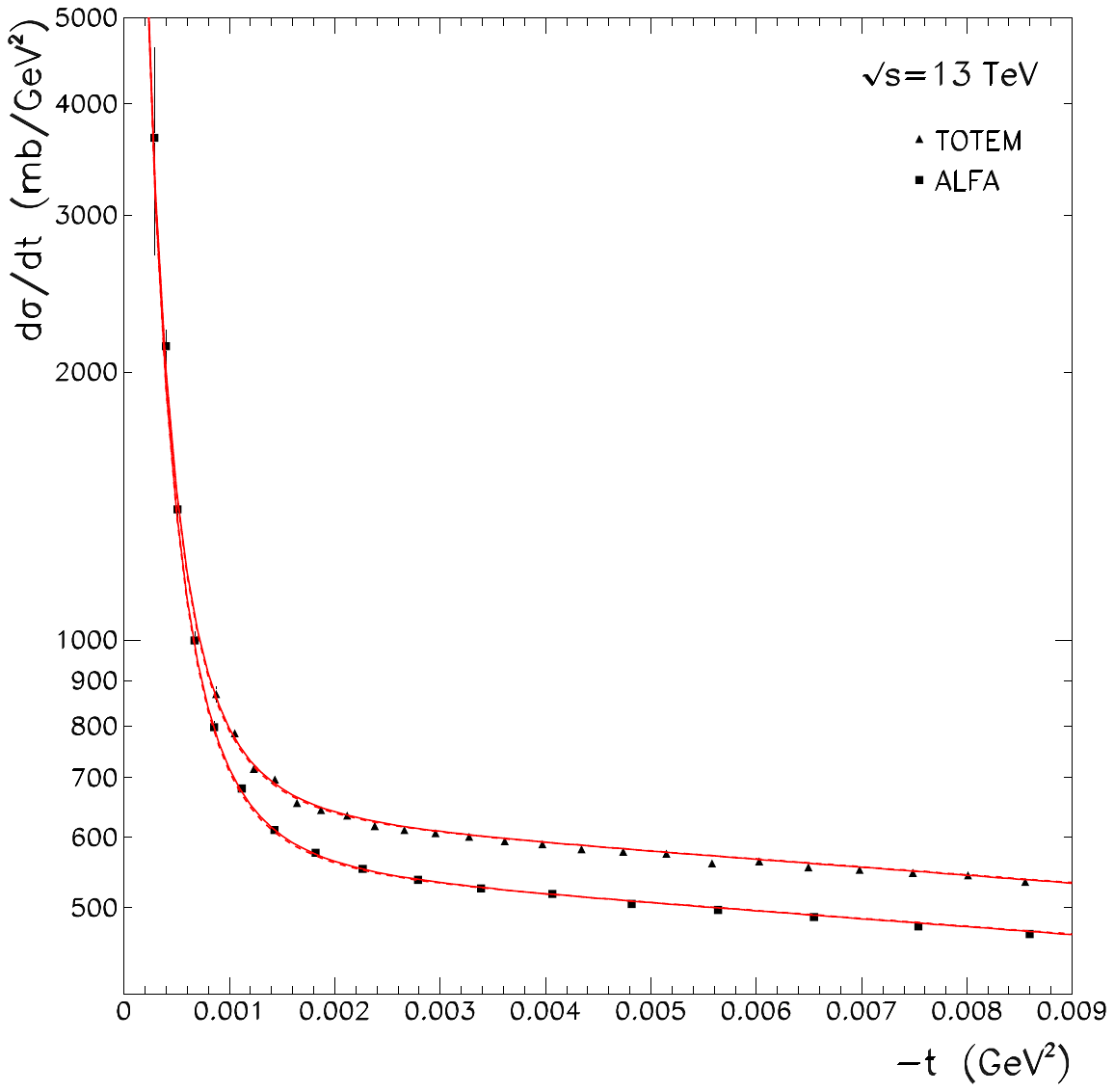}
\vspace{-1.4cm}
\caption{\small The energy behaviour of $\sigma_{tot}$ and $\sigma_{el}$ (left) and the description of $t$- dependence of elastic $pp$ differential cross section at  13 TeV in Coulomb-nuclear interference region (right). The data are from \cite{rho-T,atl13}. Theoretical curves are multiplied by the corresponding normalization factors.}
\end{center}
\end{figure*}

\begin{table*}
\centering
\caption{Values of the parameters obtained in the global fits to the Ensemble, including the TOTEM and ATLAS data. }
\vspace{0.3cm}
\begin{tabular}{c|c|c|c}
\hline\hline\\
$\beta_{\Bbb P}(0)$ &$\epsilon$ &$\alpha^{\prime}_{\Bbb P}$ (GeV$^{-2}$) &$\beta_{\Bbb O}(0)$ \\
\hline\\
2.259$\pm$0.016&0.1180$\pm$0.0020 &0.128$\pm$0.022 &
0.90$\pm$0.18\\
\hline\hline\\
$A$ (GeV$^{-2}$)&$B$ (GeV$^{-4}$&$C$ (GeV$^{-6}$)& D=A/2\\ \hline \\ 
4.78$\pm$ 0.21& 6.7$\pm$ 1.1 & 17.7$\pm$ 4.0 & \\
\hline\hline
\end{tabular}
\label{tab001}
\end{table*}

\end{document}